\documentclass[fleqn,usenatbib]{mnras}

\usepackage{newtxtext,newtxmath}

\usepackage[T1]{fontenc}

\DeclareRobustCommand{\VAN}[3]{#2}
\let\VANthebibliography\thebibliography
\def\thebibliography{\DeclareRobustCommand{\VAN}[3]{##3}\VANthebibliography}

\usepackage{graphicx}	
\usepackage{amsmath}	

\title[PUMPSIV]{Pushchino multibeams pulsar search. IV. Detection of new pulsars 
at declinations $-9^o < \delta < +55^o$.
}

\author[S. A. Tyul'bashev et al.]{
S.A. Tyul'bashev,$^{1}$\thanks{E-mail: serg@prao.ru (SAT)}
G.E. Tyul'basheva,$^{2}$
M.A. Kitaeva,$^{1}$
I.L. Ovchinnikov,$^{3}$
V.V. Oreshko,$^{1}$
S.V. Logvinenko$^{1}$
\\
$^{1}$ P.N. Lebedev Physical Institute of the Russian Academy of Sciences, Astro Space Center, Pushchino Radio Astronomy Observatory,\\
Radiotelescopnaya 1a, Moscow reg., Pushchino, 142290, Russia \\
$^{2}$ Institute of Mathematical Problems of Biology, brunch of Keldysh Institute of Applied Mathematics,\\  
Vitkevich 1, Moscow reg., Pushchino, 142290, Russia\\
$^{3}$ Skobeltsyn Institute of Nuclear Physics, Lomonosov Moscow State University, Moscow, 119234, Russia\\
}

\date{Accepted XXX. Received YYY; in original form ZZZ}

\pubyear{2015}

\begin{document}
\label{firstpage}
\pagerange{\pageref{firstpage}--\pageref{lastpage}}
\maketitle

\begin{abstract}
The search for pulsars in monitoring data obtained at the radio telescope Large Phased Array (LPA) at a frequency of 111 MHz was carried out. Daily round-the-clock observations were carried out for about 3,000 days. The duration of the observation session for each direction in the sky was 3.5 minutes per day. The search for pulsars was carried out using power spectra. To search for weak pulsars, power spectra were summed up. The expected increase in sensitivity was 35-40 times compared to observations in one session. In a blind search, 330 pulsars with regular radiation were detected, with periods (P) from 0.0333 to 3.7455 s and dispersion measures (DM) up to 249 pc/cm$^3$. 39 pulsars turned out to be new. Average profiles were obtained for 6  pulsars. The DM for 7 pulsars previously detected on the LPA have been clarified.

\end{abstract}




\section{Introduction}

Pulsars were discovered at a frequency of 81.5 MHz, which corresponds to a wavelength of 3.6 meters (\citeauthor{Hewish1968}, \citeyear{Hewish1968}). Immediately after their discovery, it became clear that the search for pulsars in the meter range is difficult for a number of reasons. First, the temperature of the Galaxy background depends on the frequency of observations and grows as $\nu^{-2.55}$ (\citeauthor{Turtle1962}, \citeyear{Turtle1962}; \citeauthor{Haslam1982}, \citeyear{Haslam1982}), whereas the pulsar flux density to low frequencies increases as $\nu^{-1.4}-\nu^{-1.6}$ (\citeauthor{Lorimer1995}, \citeyear{Lorimer1995}; \citeauthor{Bates2013}, \citeyear{Bates2013}). As a result, the lower frequency observations (the longer the wavelength) have higher background temperature and lower sensitivity, while other things remain unchanged. Second, at high frequencies, it is possible to make the signal reception band many times wider than at low frequencies. A wider reception band provides higher sensitivity. Third, scattering ($\tau$) on the interstellar medium widens the pulses. The lower frequency observations have higher scattering ($\tau\sim \nu^{-4}$, \citeauthor{Lorimer2004} (\citeyear{Lorimer2004}). Due to scattering, detection of milliseconds pulsars are difficult, and often impossible. The half-power-width of the scattered pulse of a milliseconds pulsar can be much larger than its period, and in this case the periodic signal cannot be seen. For pulsars, the broadening of a pulse due to scattering leads to a deterioration in sensitivity. So, in the ATNF catalog\footnote{https://www.atnf.csiro.au/people/pulsar/psrcat/} (\citeauthor{Manchester2005}, \citeyear{Manchester2005}) we have not been able to find a single pulsar discovered at frequencies below 200 MHz and having a DM greater than 250 pc/cm$^3$. Apparently, this is precisely due to the scattering of pulses, which leads to a loss of sensitivity during the search. In general, about half of all known pulsars are not available for observations at low frequencies. These three indicated reasons clearly show that it is easier to provide high sensitivity in the search for pulsars in the decimeter, rather than in the meter wavelength range.

Despite the obvious difficulties of searching for pulsars at frequencies below 200 MHz, three low-frequency surveys have been launched in recent years. On spaced-apart antenna arrays Low Frequency Array (LOFAR; \citeauthor{vanHaarlem2013} (\citeyear{vanHaarlem2013})) and Murchison Widefield Array (MWA; \citeauthor{Tingay2013} (\citeyear{Tingay2013})), performed on broadband dipoles, dozens of new pulsars have been found, and hundreds of known pulsars have been redetected in a blind search (\citeauthor{Sanidas2019}, \citeyear{Sanidas2019}; \citeauthor{Bhat2023}, \citeyear{Bhat2023}). In a blind search on the upgraded radio telescope Large Phased Array (LPA; \citeauthor{Shishov2016} (\citeyear{Shishov2016}), \citeauthor{Tyulbashev2016} (\citeyear{Tyulbashev2016}), being an antenna array with a filled aperture, more than 80 new pulsars and rotating radio transients (RRAT) have been discovered\footnote{https://bsa-analytics.prao.ru/en/}, as well as about 200 known pulsars (\citeauthor{Tyulbashev2022}, \citeyear{Tyulbashev2022}). In the motivation of conducting new low-frequency surveys, the authors of the papers \citeauthor{Sanidas2019} (\citeyear{Sanidas2019}); \citeauthor{Tyulbashev2022} (\citeyear{Tyulbashev2022}); \citeauthor{Bhat2023} (\citeyear{Bhat2023}) indicated the following reasons: obtaining new models of population synthesis, investigation of the nearby population of pulsars, search for pulsars with ultra-steep spectra, search for new types of pulsars.

When searching for pulsars, it is equally important to ensure high sensitivity of the telescope due to its high effective area, wide reception band, long observation sessions, and the use of new methods of data processing. In particular, in recent years it has been shown that Fast Folding Algorithm (FFA) provides higher sensitivity than Fast Fourier Transform (FFT) (\citeauthor{Morello2020}, \citeyear{Morello2020}). The practical application of FFA at Giant Metrewave Radio Telescope (GMRT) has shown that the signal-to-noise ratio (S/N) for some new pulsars is 5-10 times higher when searching using FFA than when searching using FFT (see fig.5 in \citeauthor{Singh2023} (\citeyear{Singh2023})). The use of FFT optimized for searching for pulsars with narrow pulses made it possible to increase the S/N of candidates by 2 or more times, and to discover 37 new pulsars in the archive data of the 64-meter telescope in Parkes (\citeauthor{Sengar2023}, \citeyear{Sengar2023}). Neural networks (\citeauthor{Zhu2014}, \citeyear{Zhu2014}; \citeauthor{Bethapudi2018}, \citeyear{Bethapudi2018}; \citeauthor{Wang2019}, \citeyear{Wang2019}) are used to sifting pulsar candidates.

In 2020, when searching for pulsars on the LPA radio telescope, using data with low time-frequency resolution (6 channels of 415 kHz and sampling 0.1 s), dozens of pulsar candidates were found for which it was not possible to construct average profiles and estimate DM (\citeauthor{Tyulbashev2020}, \citeyear{Tyulbashev2020}). In the power spectra of these candidates, summed up over an interval of 5 years, 2 or more harmonics were visible. 
Candidates for new pulsars were observed in one or two neighboring beams, that is, periodic signals came from the same direction in the sky. In standard ways, when the best sessions are searched for a pulsar candidate, and then an average profile is obtained for these sessions and the DM is evaluated, nothing could be done. This is due to the fact that if on the summed up power spectra the S/N of harmonics were more than 6-7, then in the recordings for individual days the S/N of harmonics were less than 0.2. That is, despite the probable detection of a new pulsar, it was neither possible to obtain its average profile nor to estimate the DM. 

A special processing program was created to check weak candidates. It represents in the figures at the same time thousands of power spectra related to the selected direction in the sky.  Each of these thousands of spectra is the sum of the power spectra for the entire observation period. When using data with high time-frequency resolution (32 channels of 78 kHz and sampling 12.5 ms), it was shown that some of the candidates are periodic interference, some are known pulsars visible in the side lobes of the LPA, some are new pulsars (\citeauthor{Tyulbashev2022}, \citeyear{Tyulbashev2022}). It was shown in this paper that when using data with high time-frequency resolution and accumulation of power spectra over 7 years, the sensitivity should reach 0.1-0.2 mJy (in the zenith direction and beyond the Galactic plane) at the frequency of 111 MHz. This sensitivity can be an order of magnitude better than the sensitivity achieved in the LOTAAS survey conducted on LOFAR at the frequency of 135 MHz (\citeauthor{Sanidas2019}, \citeyear{Sanidas2019}).

This work is based on the Pushchino Multibeams Pulsar Search (PUMPS) project, which is dedicated to the search and study of pulsars and transients detected in monitoring observations that have been going on continuously for more than 8 years. In previous works of the cycle, a new way of searching for pulse was considered (\citeauthor{Tyulbashev2022}, \citeyear{Tyulbashev2022}), a study was conducted on the variability of RRAT discovered in Pushchino (\citeauthor{Smirnova2022}, \citeyear{Smirnova2022}), investigation of the drift periods of pulsars detected in the survey when modeling the summed up power spectra (Smirnova, MNRAS submitted).

The paper is organized as follows: the observation and processing in Section 2; results in Section 3; some discussion of results and conclusion in Section 4; and four Appendixes with additional materials. 

\section{Observations and processing}
\label{Section2}

In 2012, the modernization of the LPA radio telescope was completed (\citeauthor{Shishov2016}, \citeyear{Shishov2016}). In the course of modernization on the basis of one antenna field consisting of 16,384 dipoles, it was possible to implement the possibility of creating four radio telescopes. Two of them operate on a daily basis according to independent observation programs, one is used by the engineering service to monitor the condition of the antenna, one radio telescope has not yet been implemented. In this paper we will talk about observations on the LPA3 radio telescope. This radio telescope was specially created for the Space Weather program (\citeauthor{Shishov2016}, \citeyear{Shishov2016}). It has 128 uncontrolled beams, overlapping declinations from $-9^o$ to $+55^o$. LPA is a meridian instrument. During 24 hours, each source in the specified declinations crosses the meridian and is recorded. The physical dimension of the antenna field is $187\times 384$ meters in the east-west and north-south directions. The dimension of the LPA beam is approximately $0.5^o \times 1^o$. The intersection of the meridian at half the power of the radiation pattern takes about 3.5 minutes. The field of view of 128 beams, taking into account their intersection at the level 0.4, is approximately 50 square degrees. The effective area of the LPA3 in the zenith direction is 47,000 sq.m. (0.65 of the geometric area), which makes it one of the world's largest telescopes operating in the meter wavelength range. The central receiving frequency of the radio telescope is 110.3 MHz, the full receiving band is 2.5 MHz.

128 LPA3 beams in 2014-2021 were connected to the three recorders designed for observations under the Space Weather project (\citeauthor{Shishov2016}, \citeyear{Shishov2016}). At declinations $-9^o < \delta < +21^o$ and $+21^o < \delta < +42^o$, two recorders are connected (48+48=96 beams). In these 96 beams, daily monitoring observations have been carried out for more than 8 years. The search for pulsars in the accumulated data was carried out several times\footnote{https://bsa-analytics.prao.ru/en/}. In December 2020 , 24 beams were connected to the third recorder ($+42^o < \delta < +52^o$), and in September 2021, 8 more beams ($+52^o < \delta < +55^o$) were implemented. The area with declination $+42^o < \delta < +55^o$ has not previously been processed as part of the PUMPS survey. The signal is recorded on all recorders in hourly parts. The beginning of the hour recording is controlled by the atomic frequency standard, inside the hour recording the time is controlled by a quartz oscillator. To equalize the gain in the frequency channels, a noise signal with a temperature of 2400K is applied to the input of the amplifiers 6 times a day (\citeauthor{Tyulbashev2020}, \citeyear{Tyulbashev2020})

\begin{figure*}
	\includegraphics[width=\textwidth]{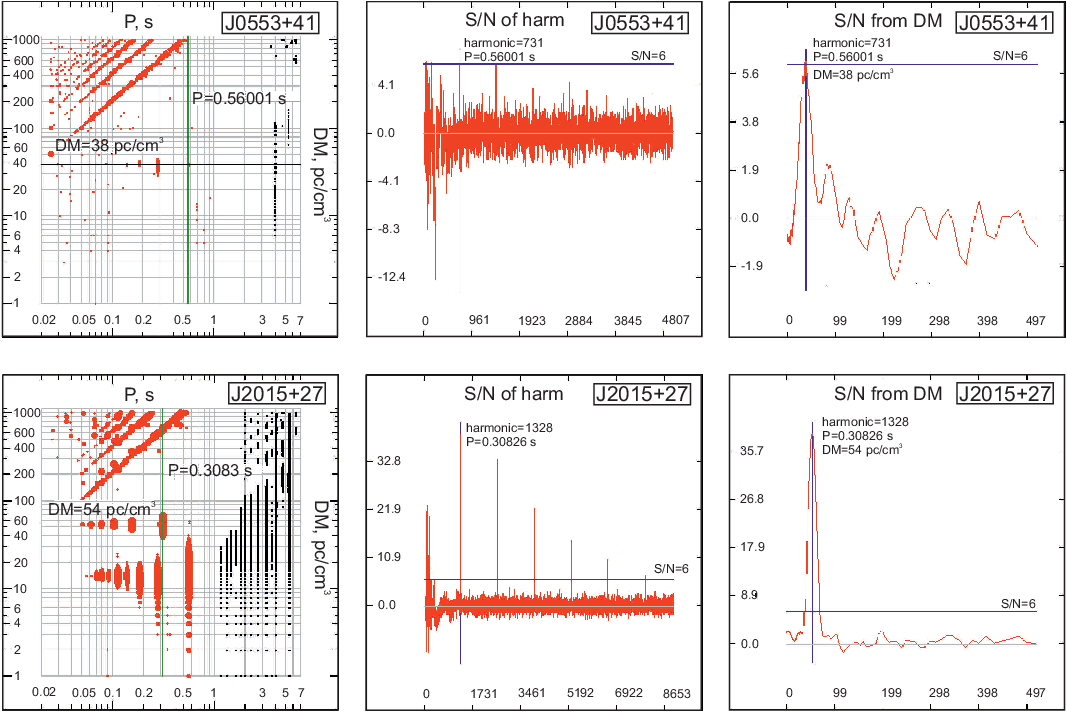}
    \caption{The maps created by the processing program show pulsars J0553+41 (upper row of figures) and J2015+27 (lower row of figures). The intersection of the horizontal and vertical lines on the left panels  indicate the DM and P of pulsars visible on the map. The middle panels show power spectra. In the left part of the power spectra, low-frequency noises are visible, which could not be removed when subtracting the baseline. The right panels represent the dependence of the harmonic height in the power spectrum on the DM being checked. The middle and right panels show S/N on vertical scales. The horizontal scales on the middle and right panels show the number of the point in the power spectrum and the DM of the pulsar. The pulsar J0553+41 has 5 harmonics visible on the power spectrum, three of them have S/N=3.5-4. The pulsar J2015+27 has 6 harmonics visible. }
    \label{fig:fig1p}
\end{figure*}

Two parallel modes of operation are implemented on all recorders. The data coming to the recorder is simultaneously recorded with low and high time-frequency resolution. As mentioned above, for data with high time-frequency resolution, the sampling for the point is 12.5 ms. This sampling is determined by the maximum velocity of the recorder. It is known that in other low-frequency surveys on the search for pulsars the sampling time is less than 1 ms (\citeauthor{Sanidas2019}, \citeyear{Sanidas2019}; \citeauthor{Bhat2023}, \citeyear{Bhat2023}). 20\% of ATNF pulsars have a pulse half-power-width ($W_e$) less than 12.5 ms. For such pulsars, the sensitivity of PUMPS will deteriorate. 
The 78 kHz channel width for the recorded data is also too large for the organization of full-fledged pulsar observations. Already at DM=50 pc/cm$^3$ dispersion smearing inside the frequency channel will smooth the data twice and worsen the observed S/N by $2^{1/2}$= 1.4 times for pulsars whose pulse duration is 12.5 ms. Therefore, the time-frequency resolution, which is called high in this paper, is obviously insufficient for the optimal search for pulsars.

The basis of the recording system is the basic module, which is a printed circuit board with electronic components installed on it. The central element in the module is an FPGA chip (field-programmable gate array). It is the amount of FPGA resources and the speed of these resources that determine the maximum possible time-frequency resolution of the recorder as a whole. Each FPGA processes data streams coming from 8 of 12-bit Analog-to-Digital Converters (ADCs) sampling at 230 MHz in parallel, with a total input stream of 2.76 TB per second. In this case, about 80\% of the FPGA capabilities are used. The output stream of data written onto hard drives from all three recorders is 4.65 GB per hour and 39.8 TB per year. A detailed description of the recorders is considered in the paper (Logvinenko, 2023 in press).

The observations data were processed on two servers. AMD EPYC 7702 processors are installed on each server (64 computing cores, 128 threads, a clock frequency of 2 GHz with the ability to operate individual cores at a frequency of up to 3.35 GHz). The Random Access Memory (RAM) capacity is 1 TB per server. The control is carried out using the operating system Debian GNU/Linux. For data storage, the Zettabyte File System (ZFS) is used, which includes the functions of disk drive management, data integrity verification and redundancy. For communication between servers, an Infiniband network with a bandwidth of 40 Gbit/s is used. To store intermediate processing data, a virtual tmpfs file system is used, which occupies RAM up to 500 GB per server.

The search for pulsars is carried out using power spectra obtained for each observation session (one observation session per day). In the power spectrum, information about the pulse phase is lost, and the pulsar period is reflected as a harmonic, always located at the same point in the array containing the spectrum. If there is a pulsar in the recording, then when summing the spectra for different days, the S/N of harmonics corresponding to its period will increase (\citeauthor{Tyulbashev2020}, \citeyear{Tyulbashev2020}).

Approximately 2\% of all raw data has strong interference. To identify these data, we check the value of the noise track in temperature units after adding the frequency channels with DM=0 pc/cm$^3$. As shown by \citeauthor{Tyulbashev2020} (\citeyear{Tyulbashev2020}), the noise tracks for different days may differ. For each direction in the sky, we rank all the noise tracks for different days by increasing their $\sigma_n$. Signals for different days are summed up as long as the growth of the S/N continues. The ``bad'' data are excluded from further consideration.

In this paper, we have consistently calculated and summed up the power spectra obtained from data with high time-frequency resolution on all recorders for the period 2014-2022. The large RAM of the new server made it possible to use a virtual file system for the calculation, which dramatically reduced the data processing time. The program processes each hour separately, sequentially going through all days of the year, creating power spectra for a certain hour, and summing them up to the corresponding summed up spectra. The cumulative spectra for each year are stored, and intermediate data (for individual sessions) are not saved. When processing one hour of observations by one recorder, 26,880 files are generated, the total volume of which is about 100 GB. Each file contains power spectra for a certain right ascension (a 1.5-minute segment of the specified hour), declination coordinates (corresponding to one of the 48 antenna beams), and a certain pulse width. Each file contains the received arrays of power spectra for all the DMs being sorted. Thus, about 200 GB of RAM is required to account for one hour of observations, since at the same time arrays of power spectra for the current day and summed up arrays are stored in memory, to which the received spectra for the current day are added each time. The summation of the spectra takes place in RAM and this radically reduces the processing time, because after processing the next day, reading and writing to the hard disk is not required. The resulting summed up spectra are recorded to the disk once every half an hour, so as not to lose them in the event of a system failure. Power spectra are calculated in multi-threaded mode when all 128 available threads are used. The calculation is paralleled by coordinates. With this algorithm, it takes about 15 hours to calculate the summed up spectra per year for one hour of right ascension for all declinations.

After the cumulative spectra for each year are calculated, we sum them up for all years. As a result, for each direction in the sky, we get about 1,800 power spectra summed up over the entire observation interval. These are power spectra for different sorted out DM (0<DM<1,000 pc/cm$^3$)  and pulse widths (12.5<$W_e$<400ms). For low DM, we used a step of 1 pc/cm$^3$, then the step increases taking into account the broadening of the pulse due to dispersion smearing. A total of 117 trials were used in the interval of the tested DM (see Fig.1 in \citeauthor{Tyulbashev2022} (\citeyear{Tyulbashev2022}).) The obtained power spectra are viewed in a specially developed visualization program.

The method of power spectra visualization was considered in the paper \citeauthor{Tyulbashev2022} (\citeyear{Tyulbashev2022}). All the spectra corresponding to one direction in the sky are represented in one drawing (map). DM is measured on the vertical axis, and P is measured on the horizontal axis. The harmonic heights of the power spectra, expressed in signal-to-noise units (S/N), are plotted on the map in the form of circles. Harmonics with S/N $\geq 4$ are selected.There are harmonics in the power spectra that are visible in all beams of LPA3. Such harmonics are recorded in a special file which have been used by visualization program. Usually, these frequency interferences are not shown on the maps and on the power spectra, but the visualization program has the ability to show maps ``as is''. The more S/N of a harmonic, the larger is the diameter of the circle. The S/N boundary can be increased by removing weak signals from the map. Circles for harmonics of different spectra can overlap each other. By clicking on the circle, it is possible to view a list of power spectra from which this circle is composed, go to viewing these spectra, and then to plot other charts that allow you to analyze the nature of the signal.

The resulting maps are viewed visually, separately for each direction in the sky. The pulsars appear on the map as a set of vertical segments, where the first segment corresponds to the pulsar period, the second to the half of the period, etc. A horizontal line passing through the midpoints of the segments indicates the DM of the pulsar. If pulsars are detected, the program allows you to get an estimate of DM and P, view raw data, create average profiles, and much more (see \citeauthor{Tyulbashev2022} (\citeyear{Tyulbashev2022}). In this paper, we assumed that a pulsar is detected if it is visible in one or two adjacent beams, and the S/N of the highest harmonic, or sum of harmonics, is more than 8.

\section{Results}

When viewing maps showing summed power spectra for different directions in the sky, 330 pulsars were detected in a blind search. Of these, 291 are known pulsars, including all pulsars previously detected on the LPA. 39 new pulsars were discovered. The list of detected known pulsars is given in Appendix.~\ref{Appendix A}. The P/DM maps of these pulsars are presented on the website\footnote{https://bsa-analytics.prao.ru/en/pulsars/known/}, and their study will be carried out in subsequent papers. In this paper, in the section ``Discussion of results'', the number of detected ATNF pulsars in PUMPS is considered.

Fig.~\ref{fig:fig1p} shows a set of maps generated by the processing program obtained for weak and strong new pulsars (J0553+41; J2015+27). The pulsars are clearly visible on P/DM maps (left panels in the figure). The harmonics of these pulsars, including those having S/N<4, are visible on the power spectra (the middle panels in the figure). The DM are determined by the maximum of dependence (S/N)/DM, where S/N reflects the height of the harmonic in the power spectrum for different DMs (right panels in the figure). For the pulsar J0553+41 the first two harmonics reach S/N=6. With incoherent summation of harmonics visible on the power spectrum, S/N grows to 10.8. On the map with J2015+27, where the intersection of the lines indicates a pulsar, another pulsar is visible, simultaneously falling into the LPA3 directional diagram. This is the well-known pulsar J2018+2839 (P=0.5579 s and DM=14.1 pc/cm$^3$) observed in the side lobe of the LPA. 

Table~\ref{tab:tab1} shows the discovered pulsars. Columns 1-3 give the name of the pulsar and its coordinates in right ascension and declination ($\alpha_{2000}$, $\delta_{2000}$). The accuracy of the right ascension coordinate for all pulsars is $\pm 1.5$~minutes. Columns 4 and 5 show P and DM of pulsars. Column 6 shows S/N value for the highest harmonic in the power spectrum, and the number of this harmonic is given in parentheses. As the processing shows, the first harmonics often fall into the region of low-frequency noise in the power spectra, and their value in units of S/N is determined incorrectly. Sometimes part of the harmonics in the power spectra is missing. We indicate the S/N value, since it indirectly indicates the pulsar flux density and can help observers when making confirmation of pulsar detection observations (\citeauthor{Tyulbashev2023}, \citeyear{Tyulbashev2023}). Based on the noise track determined by the background of the Galaxy, S/N=6 corresponds to an integral flux density of 0.2-2 mJy for pulsars at the frequency of 111 MHz and depends on the period of the pulsar, its height above the antenna at the moment of culmination, the background temperature in the direction of the pulsar, the pulsar coordinate relative to the coordinate of the fixed beam of LPA3. These estimates are very rough, but allow to roughly estimate the expected flux density when conducting observations on other radio telescopes. Work on the flux densities of all pulsars observed in the PUMPS survey is in preparation.

\begin{table*}
	\centering
	\caption{Characteristics of the new pulsars.}
	\label{tab:tab1}
	\begin{tabular}{cccccc} 
		\hline
name & $\alpha_{2000}$ & $\delta_{2000}$ & $P_0$(s) & $DM$(pc/cm$^3$) & $S/N$\\
		\hline
J0129+36 & 01$^h$29$^m$30$^s$ & 36$^o$05$^\prime$$\pm 10$ & 1.089$\pm 0.005$ & 54$\pm$2 & 10.9(1)\\
J0149+29 & 01$^h$49$^m$00$^s$ & 29$^o$10$^\prime$$\pm 15$ & 2.654$\pm 0.010$ & 34.5$\pm$2 & 10.6(6)\\
J0553+41 & 05$^h$53$^m$30$^s$ & 41$^o$20$^\prime$$\pm 15$ & 0.5600$\pm 0.001$ & 38$\pm$2 & 6.4(2)\\
J0734+14 & 07$^h$34$^m$30$^s$ & 14$^o$43$^\prime$$\pm 15$ & 1.772$\pm 0.005$ & 50$\pm$4 & 8.2(4)\\
J1003+41 & 10$^h$03$^m$30$^s$ & 41$^o$03$^\prime$$\pm 15$ & 0.8804$\pm 0.003$ & 30$\pm$3 & 6.8(1)\\
J1105+37 & 11$^h$05$^m$00$^s$ & 37$^o$41$^\prime$$\pm 15$ & 0.8747$\pm 0.003$ & 36$\pm$3 & 9.3(3)\\
J1409+49 & 14$^h$09$^m$30$^s$ & 49$^o$44$^\prime$$\pm 15$ & 1.125$\pm 0.005$ & 29$\pm$2 & 7.5(1)\\
J1444+18 & 14$^h$44$^m$30$^s$ & 18$^o$10$^\prime$$\pm 10$ & 1.1326$\pm 0.0005$ & 16.5$\pm$1.05 & 13.5(1)\\
J1515+20 & 15$^h$15$^m$00$^s$ & 20$^o$51$^\prime$$\pm 15$ & 1.147$\pm 0.005$ & 21.5$\pm$2 & 11.8(1)\\
J1730+13 & 17$^h$30$^m$45$^s$ & 13$^o$05$^\prime$$\pm 15$ & 1.150$\pm 0.005$ & 32$\pm$2 & 27.0(1)\\
J1742+20 & 17$^h$42$^m$45$^s$ & 20$^o$17$^\prime$$\pm 15$ & 0.2526$\pm 0.0001$ & 20$\pm$2 & 8.2(2)\\
J1754+48 & 17$^h$54$^m$00$^s$ & 48$^o$46$^\prime$$\pm 15$ & 0.6519$\pm 0.002$ & 23$\pm$2 & 16.1(1)\\
J1803+47 & 18$^h$03$^m$00$^s$ & 47$^o$08$^\prime$$\pm 15$ & 0.3469$\pm 0.0010$ & 29.5$\pm$1.5 & 11.8(1)\\
J1837+10 & 18$^h$37$^m$00$^s$ & 10$^o$46$^\prime$$\pm 15$ & 1.707$\pm 0.005$ & 54$\pm$4 & 7.4(8)\\
J1848+26 & 18$^h$48$^m$15$^s$ & 26$^o$29$^\prime$$\pm 15$ & 0.6288$\pm 0.002$ & 64$\pm$2 & 8.2(1)\\
J1854+40 & 18$^h$54$^m$00$^s$ & 40$^o$05$^\prime$$\pm 15$ & 1.748$\pm 0.005$ & 37$\pm$3 & 8.7(4)\\
J1916+38 & 19$^h$16$^m$30$^s$ & 38$^o$48$^\prime$$\pm 15$ & 0.5136$\pm 0.002$ & 77$\pm$4 & 11.4(1)\\
J1918+26 & 19$^h$18$^m$30$^s$ & 26$^o$55$^\prime$$\pm 15$ & 0.7233$\pm 0.003$ & 113$\pm$4 & 8.3(1)\\
J1937+34 & 19$^h$37$^m$30$^s$ & 34$^o$04$^\prime$$\pm 15$ & 1.749$\pm 0.005$ & 98$\pm$5 & 7.1(1)\\
J1940+14 & 19$^h$40$^m$15$^s$ & 14$^o$37$^\prime$$\pm 15$ & 1.279$\pm 0.005$ & 70$\pm$3 & 9.5(3)\\
J1948+06 & 19$^h$48$^m$45$^s$ & 06$^o$02$^\prime$$\pm 15$ & 1.325$\pm 0.005$ & 61$\pm$3 & 11.3(2)\\
J2000+22 & 20$^h$00$^m$00$^s$ & 22$^o$13$^\prime$$\pm 10$ & 1.0505$\pm 0.0005$ & 86$\pm$3 & 52.4(1)\\
J2000+04 & 20$^h$00$^m$15$^s$ & 04$^o$07$^\prime$$\pm 15$ & 0.7768$\pm 0.003$ & 60$\pm$6 & 7.3(1)\\
J2014+10 & 20$^h$14$^m$15$^s$ & 10$^o$28$^\prime$$\pm 10$ & 1.140$\pm 0.005$ & 74$\pm$4 & 7.0(2)\\
J2015+27 & 20$^h$15$^m$15$^s$ & 27$^o$48$^\prime$$\pm 10$ & 0.3081$\pm 0.0001$ & 55$\pm$2 & 54.2(1)\\
J2031+34 & 20$^h$31$^m$00$^s$ & 34$^o$19$^\prime$$\pm 10$ & 1.819$\pm 0.005$ & 105$\pm$4 & 24.3(1)\\
J2040+20 & 20$^h$40$^m$00$^s$ & 20$^o$42$^\prime$$\pm 15$ & 0.2905$\pm 0.0005$ & 31$\pm$2 & 13.6(2)\\
J2043+31 & 20$^h$43$^m$00$^s$ & 31$^o$27$^\prime$$\pm 15$ & 0.9368$\pm 0.003$ & 145$\pm$5 & 17.7(1)\\
J2116+37 & 21$^h$16$^m$30$^s$ & 37$^o$04$^\prime$$\pm 15$ & 0.1459$\pm 0.0005$ & 43$\pm$2 & 8.4(1)\\
J2124+34 & 21$^h$24$^m$00$^s$ & 34$^o$27$^\prime$$\pm 15$ & 0.4891$\pm 0.002$ & 84.5$\pm$2 & 15.5(1)\\
J2151+19 & 21$^h$51$^m$15$^s$ & 19$^o$11$^\prime$$\pm 15$ & 1.036$\pm 0.005$ & 31$\pm$2 & 10.7(5)\\
J2201+33 & 22$^h$01$^m$15$^s$ & 33$^o$00$^\prime$$\pm 10$ & 0.9655$\pm 0.003$ & 78$\pm$2 & 30.1(1)\\
J2226-03 & 22$^h$26$^m$15$^s$ & -03$^o$15$^\prime$$\pm 20$ & 0.7697$\pm 0.0002$ & 17.5$\pm$1.5 & 18.0(5)\\
J2229+40 & 22$^h$29$^m$00$^s$ & 40$^o$26$^\prime$$\pm 10$ & 0.2729$\pm 0.0010$ & 74$\pm$2 & 42.5(1)\\
J2300+52 & 23$^h$00$^m$30$^s$ & 52$^o$20$^\prime$$\pm 15$ & 0.4265$\pm 0.0010$ & 82$\pm$4 & 7.9(1)\\
J2302+48 & 23$^h$02$^m$45$^s$ & 48$^o$14$^\prime$$\pm 15$ & 0.7421$\pm 0.0002$ & 72$\pm$4 & 16.8(2)\\
J2305+19 & 23$^h$05$^m$30$^s$ & 19$^o$40$^\prime$$\pm 15$ & 0.2693$\pm 0.0010$ & 20.5$\pm$1.5 & 12.9(1)\\
J2312+21 & 23$^h$12$^m$30$^s$ & 21$^o$40$^\prime$$\pm 15$ & 1.256$\pm 0.005$ & 19$\pm$3 & 9.3(2)\\
J2355+04 & 23$^h$55$^m$30$^s$ & 04$^o$43$^\prime$$\pm 15$ & 0.9576$\pm 0.003$ & 13$\pm$2 & 7.2(2)\\
		\hline
	\end{tabular}
\\
Note: The accuracy of the right ascension coordinate for all pulsars is ±1.5 minutes.
\end{table*}

Comments on some of the  detected pulsars:\\
J0227+34: 02$^h$27$^m$00$^s$; 34$^o$01$^\prime$$\pm 15^\prime$; P=1.2443$\pm 0.005$~s; DM=27$\pm$2 pc/cm$^3$. When preparing the publication for printing, it were found the paper (\citeauthor{Dong2023}, \citeyear{Dong2023}) with a published pulsar J0227+3356, having P=1.2401 and DM=27 pc/cm$^3$. Within the limits of coordinate, P and DM errors, the sources are the identical. Apparently, we independently discovered the same pulsar;\\
J0628+07: 06$^h$27$^m$00$^s$; 07$^o$08$^\prime$$\pm 10^\prime$; P=0.2381$\pm 0.0005$~s; DM=138$\pm$3 pc/cm$^3$. In the paper \citeauthor{Chandler2003} (\citeyear{Chandler2003}), The pulsar J0627+0706 is indicated, having DM=138.25 pc/cm$^3$, P=0.4758 s. Its coordinates and DM are close to those obtained for the pulsar we discovered. The period of J0628+07 we estimate from the power spectrum as 0.2381 s, and it is half as small as in the paper \citeauthor{Chandler2003} (\citeyear{Chandler2003}). We assume that J0628+07 and J0627+0706 are the same object whose period needs independent verification;\\
J1556+00: 15$^h$56$^m$00$^s$; 00$^o$52$^\prime$$\pm 15^\prime$; P=0.5768$\pm 0.0002$~s; DM=19.5$\pm$1.5 pc/cm$^3$. In the paper \citeauthor{Tyulbashev2018AZh} (\citeyear{Tyulbashev2018AZh}), there is RRAT J1555+01, having DM=16-20 pc/cm$^3$. The closeness of the coordinates and the dispersion measures of the new pulsar and the transient suggests that, most likely, the periodic radiation of RRAT has been discovered;\\
J1848+15. This object is missing in Table~\ref{tab:tab1}. However, in the paper \citeauthor{Tyulbashev2018AA} (\citeyear{Tyulbashev2018AA}) the pulsar was discovered as RRAT. In the paper \citeauthor{Sanidas2019} (\citeyear{Sanidas2019}) it is noted that the pulsar was found by single pulses, but its regular radiation was also found. In the present paper, no periodic radiation was detected during the accumulation of power spectra over 8 years of observations;\\
J2038+35: 20$^h$38$^m$30$^s$; 35$^o$19$^\prime$$\pm 15^\prime$; P=0.0801$\pm 0.0002$~s; DM=58$\pm$1 pc/cm$^3$. In the paper \citeauthor{Hessels2008} (\citeyear{Hessels2008}) it says about the discovery of a pulsar J2038+35 with P=0.16 s and DM=57.9 pc/cm$^3$. The periods differ twice, but the dispersion measures and coordinates are close. Most likely, the same object has been discovered, the period of which needs independent verification;\\
J2048+12: 20$^h$48$^m$00$^s$; 12$^o$58$^\prime$$\pm 15^\prime$; P=2.9168$\pm 0.0005$~s; DM=36$\pm$2 pc/cm$^3$. In the paper \citeauthor{Logvinenko2020} (\citeyear{Logvinenko2020}) it was said about the discovery of RRAT J2047+12, having DM=36$\pm 2$ pc/cm$^3$. The closeness of the coordinates and the DM indicates the probable detection of RRAT periodic radiation;\\
J2203+21: 22$^h$03$^m$00$^s$; 21$^o$34$^\prime$$\pm 15^\prime$; P=1.365$\pm 0.005$~s; DM=17$\pm$1 pc/cm$^3$. In the paper \citeauthor{Tyulbashev2018AZh} (\citeyear{Tyulbashev2018AZh}) it was said about the discovery of RRAT J2202+21, having DM=17$\pm 2$ pc/cm$^3$. Apparently, we have detected regular radiation of this transient.

In total, in this work, regular radiation was detected of 8 pulsars discovered on the LPA as RRAT (\citeauthor{Tyulbashev2018AZh}, \citeyear{Tyulbashev2018AZh}; \citeauthor{Tyulbashev2018AA}, \citeyear{Tyulbashev2018AA}): J0317+13; J1404+11, J1556+00; J2048+12; J2051+12; J2105+19; J2203+21; J2209+22. For 7 pulsars detected in PUMPS earlier, it was possible to refine the DM estimates (see Table~\ref{tab:tab2}). Except for the pulsar J1921+34, the previous DM estimates, within the error limits, coincide with the new values. In the first column of the table, the pulsar's name is given according to the ATNF catalog, in the second and third columns, the previous and updated values of the DM are given.

\begin{table}
	\centering
	\caption{Refined DM values.}
	\label{tab:tab2}
	\begin{tabular}{ccc} 
		\hline
name & $DM_{ATNF}$(pc/cm$^3$) & $DM$(pc/cm$^3$)\\
		\hline
J0220+36 & 40$\pm$10 & 45$\pm$2\\
J1536+17 & 25$\pm$6 & 29$\pm$2\\
J1832+27 & 46$\pm$3 & 48$\pm$2\\
J1917+17 & 41$\pm$3 & 39$\pm$2\\
J1921+34 & 80$\pm$3 & 90$\pm$2\\
J2105+19 & 35$\pm$5 & 34.5$\pm$1.5\\
J2333+20 & 17$\pm$5 & 12$\pm$2\\
		\hline
	\end{tabular}
\end{table}

\begin{figure*}
	\includegraphics[width=\textwidth]{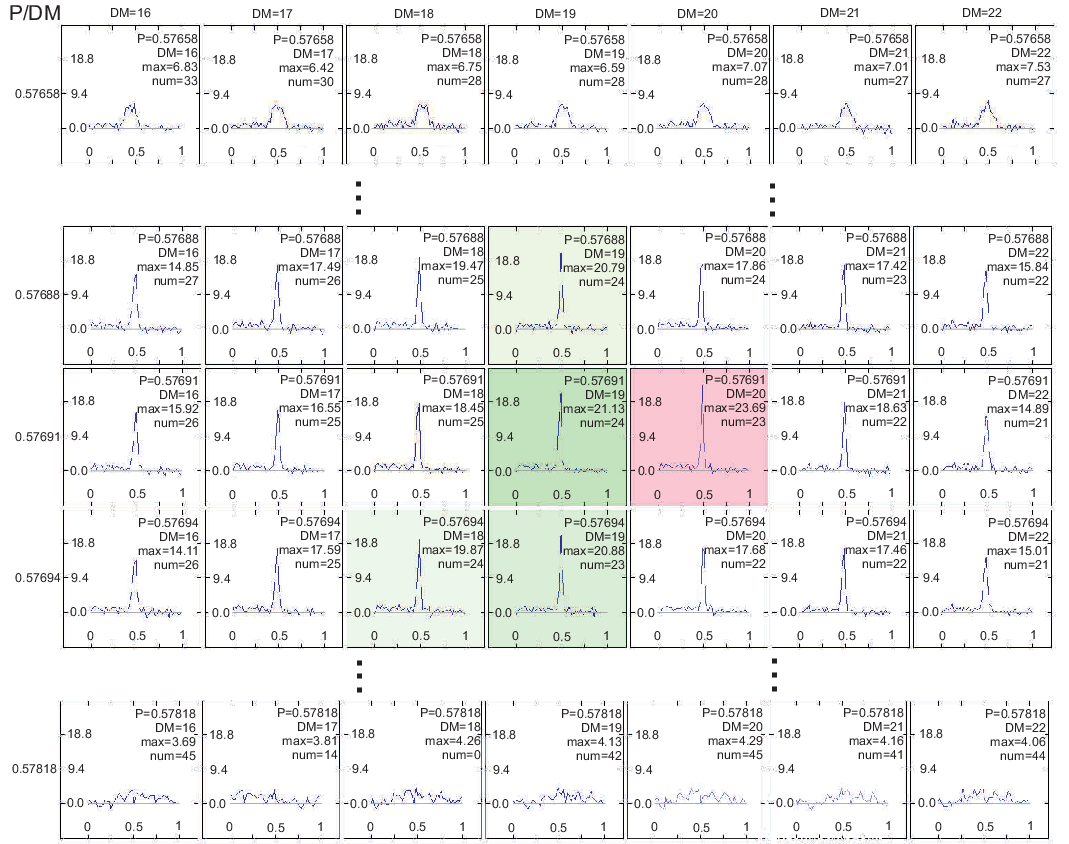}
    \caption{For pulsar J1556+00, three fragments of average profiles with different P and DM from the set of all profiles constructed by the visualization program are given. The tested DM is shown horizontally on the top, the tested P is shown vertically on the left. Inside the picture of the average profile, the P and DM being checked are shown from top to bottom, the value of the maximum in the profile in units of S/N (max), the number of the point in the profile where the visible maximum is (num). When searching for the summed power spectra for J1556+00, the values P=0.5774 s, DM=22$\pm$4 pc/cm$^3$ were obtained. When obtaining the average profiles, the periods were sorted out from 0.57658 s to 0.57818 s (upper and lower panels), the DM was sorted out from 16 pc/cm$^3$  to 22 pc/cm$^3$  (extreme left and right columns).When checking the observation sessions, it was found out that the maxima of the average profiles are concentrated around P=0.5769 s and DM=19.5$\pm$1.5 pc/cm$^3$. On the middle panel of the drawing in the central part, colored medium profiles are visible, indicating the detection of a pulsar. The upper panel shows a profile that has blurred due to an incorrect period (P = 0.57658 s). On the bottom panel, when checking the period P=0.57818 s, the profile completely disappeared.}
    \label{fig:fig2p}
\end{figure*}

In addition to P and DM, the average pulsar profile is one of the basic characteristics. With the summation of several hundred, or better, several thousand periods, the shape of the average profile becomes stable and can be used to construct radiation models (\citeauthor{Lorimer2004}, \citeyear{Lorimer2004}). However, in the case of searching for pulsars using summed up power spectra, obtaining the average pulsar profile in the LPA data for one observation session may not be a solvable task. To obtain an average profile, it is necessary to find a session during which the received signal was strong enough to see it. For example, for pulsars located at declinations -9$^o$ < $\delta$ <+42$^o$, observations were used at an interval of 8.3 years (about 3,000 sessions for each direction). According to the paper \citeauthor{Tyulbashev2022} (\citeyear{Tyulbashev2022}) S/N increases in the summed up power spectrum by about 36 times over 8 years. For the strongest of pulsar found (J2015+27), S/N=54.2 for the highest harmonic. For this pulsar in individual sessions, the average harmonic height should be S/N=54.2/36$\approx$1.5. It means that when P and DM are iterated over in sessions lasting 3.5 minutes, the S/N of a pulse in the average profile may be too low to be sure of obtaining a profile. So in the paper \citeauthor{Tyulbashev2022} (\citeyear{Tyulbashev2022}), using the same search as in the present study, out of 11 new pulsars, the average profile was obtained for only one pulsar. At the same time, pulsars are sources with variable radiation. Due to the intrinsic variability of pulsars and the variability due to interstellar scintillations, it is possible that on some days it is possible to obtain an average pulsar profile, even if the pulsar is weak in general.

\begin{figure*}
	\includegraphics[width=\textwidth]{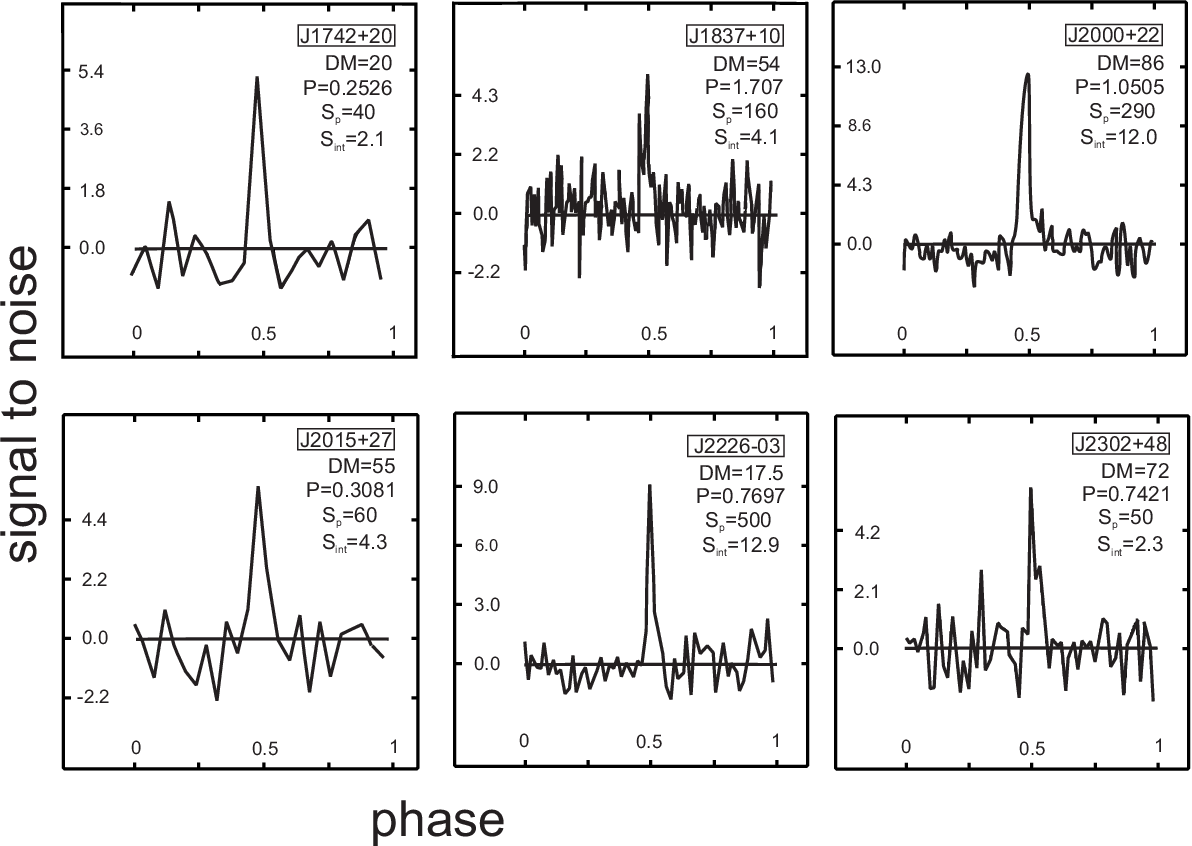}
    \caption{The figure shows the average profiles of pulsars. In the upper right corner of the pictures are the names of pulsars, P, DM and $S_p$. The P and DM are taken from Table~\ref{tab:tab1}. The axes in the figure are the same as in Figure 2.}
    \label{fig:fig3pn}
\end{figure*}

The pulsar's P and DM estimates obtained from the power spectrum have errors. To obtain the average pulsar profile, it is necessary to iterate over the values of P and DM within the error limits. We have conducted such a search for all new pulsars. For each tested value of P and DM, an average profile per observation session was obtained. In total, when checking one session, more than 400 average pulsar profiles were obtained. The number of the point with the maximum value (in S/N units) inside the resulting profile was stored, and an annular shift was performed until the maximum point was in the middle of the profile. The horizontal scale of the profile is shown in the figure in phase values, so the middle of the profile corresponds to phase 0.5. It is possible to display simultaneously  all the drawings of average profiles per session, or any individual profile. For a real average profile in the general drawing with all average profiles, the maximum S/N should be in the pictures with close P and DM. For the convenience of viewing a large number of images with average profiles, the image with the maximum value of S/N is colored pink by the visualization program, the picture with the next value is colored green, and the pictures with the next three values of the height of the average profile are colored light green. Thus, the concentration of multi-colored profile pictures in one place indirectly indicates that the real average profile has been obtained.

The program has the possibility to sort sessions by the maximum value of S/N of the average profiles obtained in it, and we display pictures of average profiles in descending order of S/N in sessions for different days. Therefore, to verify the existence or absence of an average profile in about 3,000 available sessions, it is enough to view only a few dozen days with the best S/N.

If the maximum value in the average profile of S/N is >6-7, then there is usually no doubt about the correctness of obtaining the profile. In a true average profile, color images concentrate on close values of P and DM, details can be seen on the profile. 
The more P and DM of variance differ from their true values, the lower the height of the average profile. At some distance from the P and DM values, the average profiles may disappear altogether. If the height of the average profile is low (S/N<6), then indirect signs are checked. The half-power-width of the profile cannot be less than the values from the expected scattering and dispersion smearing in the frequency channels. The color images should be concentrated near certain values of P and DM. The phases of the profile maxima for close dispersion measures should coincide, it is easy to see by the numbers of the points of the profile maxima, they should be the same. The visualization program also allows to output profiles ``as is'' (without shifting the maximum to the phase value 0.5) – then the phase inside the observation session can be seen with eyes. Average profiles with maxima S/N<4 were not considered.

For pulsars from Table~\ref{tab:tab1}, we have created average profiles for all observational sessions. Fig.~\ref{fig:fig2p} shows a fragment of the map generated by the visualization program.

In total, it was possible to obtain average profiles for 6 pulsars out of 39 checked. Fig.~\ref{fig:fig3pn} shows these average profiles. Based on the background temperature in the direction of the pulsar, the known effective area and the S/N estimate, the peak and integral flux densities in mJy ($S_p, S_{int}$) was estimated. Since thousands of observation sessions were searched  to extract these profiles and profiles with a maximum S/N were searched, the $S_p$ should be close to the maximum emission of these pulsars at a frequency of 111 MHz.

A blind search revealed 291 known pulsars. Of them in places with coordinates -9$^o$<$\delta$<+21$^o$ and  +21$^o$<$\delta$<+42$^o$ 69 pulsars were added, which were not detected in early processing using 6-channel data. Area with declination +42$^o$<$\delta$<+55$^o$  has not been processed before, 39 known pulsars were found in it. Maps showing the detection of known pulsars are posted on the website \footnote{https://bsa-analytics.prao.ru/en/pulsars/known/}. The area with declination $\delta$>+55$^o$ is currently being processed. Observations on this area are carried out on the LPA1 radio telescope in 8-beam mode. Data is recorded in 128 frequency channels, with a channel width of 19.5 kHz and a sampling of 3.1 ms. After the search is completed on LPA1, the list of known pulsars detected in the blind search will be increased, and an analysis will be carried out for all pulsars.

\section{Discussion of the results and conclusion}

According to ATNF data for May 2023, the total number of pulsars (P>0.025c, DM<200 pc/cm$^3$, -9$^o$<$\delta$<+55$^o$) that can be detected in PUMPS is 740. However, only 330 pulsars were found in a blind search. At the same time, we declare (\citeauthor{Tyulbashev2022}, \citeyear{Tyulbashev2022}), that we have one of the highest sensitivity in the world when searching for seconds pulsars in the meter wavelength range. The obtained search results seem strange. Below we excluded pulsars $+42^o<\delta < +55^o$ from the analysis. In this area, the total accumulation time was small, so the sensitivity was low.

\begin{figure*}
	\includegraphics[width=\textwidth]{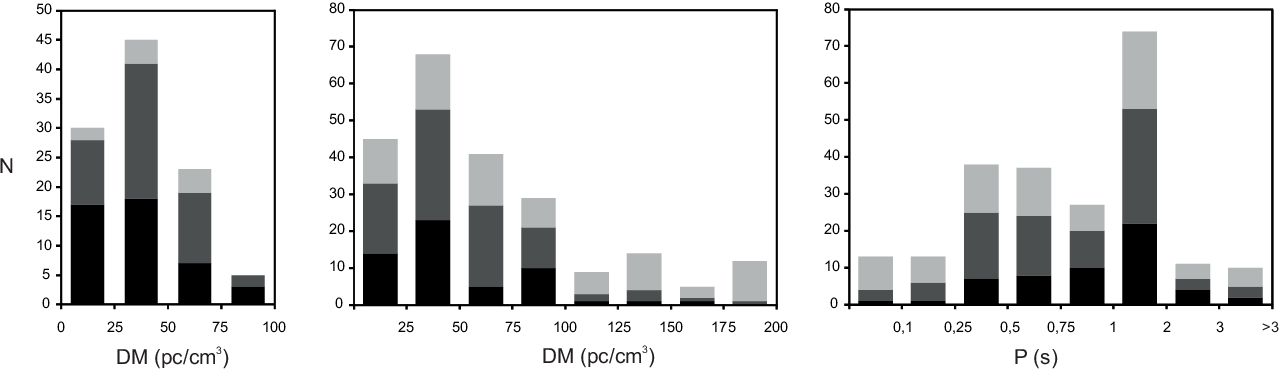}
    \caption{The left and middle panels represent columns 2-4 and 8-10 from Table~\ref{tab:tab3}, the right panel represents columns 5-7 from Table~\ref{tab:tab4}. The vertical scale shows the total number of pulsars (the bottom line in Table~\ref{tab:tab3} and Table~\ref{tab:tab4}). On the horizontal scale, the DM boundaries (left and middle panels) and the period boundaries (right panel) are shown.}
   \label{fig:fig2_2}
\end{figure*}

Sensitivity calculations in the paper \citeauthor{Tyulbashev2022} (\citeyear{Tyulbashev2022}) shown that the sensitivity in PUMPS should be maximum only when searching for pulsars with P>0.5s and DM<100 pc/cm$^3$. If P<0.5s and DM>100 pc/cm$^3$, the sensitivity drops sharply and there may be a fraction of known pulsars that are not detected in a blind search. Nevertheless, if we talk about known pulsars, in a blind search, pulsars were found J1951+3001 (P=2.7889s; DM=249 pc/cm$^3$) and J0534+2200 (P=0.0333s; DM= 56.7 pc/cm$^3$). For new pulsars, the minimum period was observed for the pulsar J2116+37 (P=0.1459s; DM=43 pc/cm$^3$), and a maximum DM – for pulsar J2043+31 (P=0.9368s; DM=145 pc/cm$^3$). 

The declared sensitivity of PUMPS is at par with the sensitivity of all earlier surveys conducted in the meter wave range by an order of magnitude or more (\citeauthor{Tyulbashev2022}, \citeyear{Tyulbashev2022}). If the sensitivity is really such, all known pulsars should be detected in the survey. In 2009  (\citeauthor{vanLeeuwen2010}, \citeyear{vanLeeuwen2010}) estimates were made that showed the planned discovery of 900 pulsars in the LOTAAS survey conducted on the LOFAR radio telescope. Taking into account the higher sensitivity of PUMPS and the comparable viewing area (see also arguments in (\citeauthor{Tyulbashev2022}, \citeyear{Tyulbashev2022}), the expected number of pulsar discoveries on LPA3 should be more than 900.

Let us analyze the pulsars detected in the blind search that fall into the areas -9$^o$<$\delta$<+42$^o$ (Table~\ref{tab:tab3} and Table~\ref{tab:tab4}). Table~\ref{tab:tab3} shows the number of pulsars for different DM. In the first column of the table, DM is given. Columns 2-4, 5-7, 8-10, 11-13 show the number of pulsars selected with different criteria for DM: DM < 100 pc/cm$^3$,  P>0.5 s, +21$^o$<$\delta$< +42$^o$, |b|>10 (pulsars outside the Galactic plane); 
DM < 200 pc/cm$^3$,  P>0.5 s, +21o <$\delta$< +42o; DM < 200 pc/cm$^3$, P>0.025 s, +21$^o$ <$\delta$< +42$^o$; DM<200 pc/cm$^3$,  P>0.025 s,   -9$^o$ <$\delta$< +42$^o$. Columns 2, 3, 4 (and similarly in the following columns) show the number of pulsars discovered in PUMPS with regular radio emission, including the present work, the number of pulsars detected in PUMPS from the ATNF catalog, the number of pulsars not detected in PUMPS from the ATNF catalog. The last row shows the number of pulsars in the column. The data for the tables were taken from the ATNF catalog and on our website\footnote{https://bsa-analytics.prao.ru/en/}. 

\begin{table*}
	\centering
	\caption{Number of detected and not detected pulsars in PUMPS depending on their DM}
	\label{tab:tab3}
	\begin{tabular}{ccccccccccccc} 
		\hline
$DM/N$ & 2 & 3 & 4 & 5 & 6 & 7 & 8 & 9 & 10 & 11 & 12 & 13\\
		\hline
0-25 & 17 & 11 & 2 & 17 & 14 & 5 & 14 & 19 & 12 & 24 & 41 & 35\\
26-50 & 18 & 23 & 4 & 22 & 27 & 4 & 23 & 30 & 15 & 38 & 58 & 64\\
51-75 & 7 & 12 & 4 & 9 & 18 & 7 & 5 & 22 & 14 & 10 & 43 & 64\\
76-100 & 3 & 2 & 0 & 9 & 7 & 4 & 10 & 11 & 8 & 12 & 24 & 61\\
101-125 & 0 & 0 & 0 & 1 & 3 & 4 & 1 & 2 & 6 & 1 & 3 & 42\\
126-150 & 0 & 0 & 0 & 1 & 3 & 5 & 1 & 3 & 10 & 2 & 4 & 46\\
151-175 & 0 & 0 & 0 & 1 & 1 & 1 & 1 & 1 & 3 & 1 & 5 & 23\\
176-200 & 0 & 0 & 0 & 0 & 1 & 9 & 0 & 1 & 11 & 0 & 1 & 60\\
		\hline
total & 45 & 48 & 10 & 60 & 74 & 39 & 55 & 89 & 79 & 88 & 179 & 395\\
		\hline
	\end{tabular}
\end{table*}

Table~\ref{tab:tab4} shows the number of pulsars entering the viewing area, depending on their periods.The first column of Table~\ref{tab:tab4} shows the boundaries of pulsar periods. Columns 2-4, 5-7 and 8-10 show the number of pulsars selected by periods with different criteria: DM < 200 pc/cm$^3$, P>0.5 s, +21$^o$<$\delta$<+42$^o$; DM < 200 pc/cm$^3$,  P>0.025 s, +21$^o$<$\delta$<+42$^o$; DM < 200 pc/cm$^3$,  P>0.025 s, -9$^o$ <$\delta$<+42$^o$ Criteria for selecting pulsars for columns 2-4, 5-7 and 8-10 in Table~\ref{tab:tab4} repeat the criteria for columns 5-7, 8-10, 11-13 in Table~\ref{tab:tab3}.

\begin{table}
	\centering
	\caption{Number of detected and not detected pulsars in PUMPS depending on their P}
	\label{tab:tab4}
	\begin{tabular}{cccccccccc} 
		\hline
$P/N$ & 2 & 3 & 4 & 5 & 6 & 7 & 8 & 9 & 10\\
		\hline
0.025-0.1 & - & - & - & 1 & 3 & 9 & 1 & 5 & 42\\
0.1-0.25 & - & - & - & 1 & 5 & 7 & 4 & 8 & 27\\
0.25-0.5 & - & - & - & 7 & 18 & 13 & 10 & 51 & 79\\
0.5-0.75 & 10 & 17 & 13 & 8 & 16 & 13 & 11 & 27 & 57\\
0.75-1 & 16 & 12 & 6 & 10 & 10 & 7 & 20 & 24 & 51\\
1-2 & 28 & 37 & 14 & 22 & 31 & 21 & 33 & 53 & 91\\
2-3 & 5 & 3 & 4 & 4 & 3 & 4 & 6 & 8 & 32\\
>3 & 1 & 5 & 3 & 2 & 3 & 5 & 3 & 3 & 16\\
  \hline
total & 60 & 74 & 39 & 55 & 89 & 79 & 88 & 179 & 395\\
		\hline
	\end{tabular}
\end{table}

In addition to the above, the distribution of new pulsars in the plane and outside the plane of the Galaxy differs from the selection of ATNF pulsars. So in the Galaxy plane |b|<10$^o$, there are 422 (57\%) of ATNF pulsars. 89  pulsars were detected in PUMPS, 22 of them in the Galactic plane (25\%). For the RRATs detected on LPA3, approximately the same ratio is observed: out of the 48 transients detected, 11 are in the plane of the Galaxy (23\%).

During the survey and taking into account this study, about 130 pulsars and RRAT were discovered, which is 7 times less than the expected discovery of more than 900 new pulsars. That is, there is a large shortage of pulsars in the meter range. The highest sensitivity and the lowest amount of interference in PUMPS corresponds to the area with declinations +21$^o$<$\delta$<+42$^o$ after cutting out the plane of the Galaxy (|b|>10$^o$) and for pulsars with periods P>0.5 s (\citeauthor{Tyulbashev2022}, \citeyear{Tyulbashev2022}). The histogram (left panel Fig.~\ref{fig:fig2_2}) shows all pulsars from this area. The height of the column shows the total number of pulsars. Pulsars discovered in PUMPS are shown in black, pulsars from the ATNF catalog found by blind search in the survey are shown in dark gray, ATNF pulsars that could not be detected are shown in light gray. Out of 103 pulsars that got into the area, 93 (90.3\%) were detected. Half of the detected pulsars were discovered on LPA3. Taking into account the detected RRAT, which are not shown on the histogram, more than half of all pulsars in the area were discovered on the LPA. For comparison, let us mention the work done on the 64-meter telescope in Parks (\citeauthor{Manchester1996}, \citeyear{Manchester1996}), according to which 94.2\% of known pulsars were detected in a blind search, which is close to our values.

When the Galactic plane is included, pulsars with DM>100 pc/cm$^3$, appear and the percentage of detected known pulsars decreases. At DM<100 pc/cm$^3$, 86\% of known pulsars were detected. For the detected pulsars, it turned out again that about half of them are discovered in PUMPS, and half are ATNF pulsars. Less than half (36.7\%) of known pulsars were detected on 100<DM<200 pc/cm$^3$.  Of the detected pulsars, the number of pulsars discovered in PUMPS dropped to a third.

Columns 8-10 of the Table~\ref{tab:tab3} and the middle panel of Fig.~\ref{fig:fig2_2} show that if pulsars with periods up to 0.025 s are added to the sample, the percentage of detected pulsars drops again in the +21$^o$<$\delta$<+42$^o$ area. At DM<100 pc/cm$^3$, 73.2\% of known pulsars are visible in PUMPS. At examples of dispersion of 100<DM<200 pc/cm$^3$, the number of discovered pulsars drops to 25\%. When checking the pulsar periods, columns 5-7 in Table~\ref{tab:tab4} (right panel Fig.~\ref{fig:fig2_2}), it can be seen that the largest percentage of pulsar losses falls on periods P<0.25 s and P>3 s.

When adding an area -9$^o$<$\delta$<+21$^o$, where a lot of interference is recorded, the number of ATNF pulsars detected by us decreases sharply. At DM<100 pc/cm$^3$ only half (52.7\%) of known pulsars were detected. At DM>100 pc/cm$^3$, only 9\% of known pulsars were detected.

Thus, Table~\ref{tab:tab3} and Table~\ref{tab:tab4}, as well as the histograms in Fig.~\ref{fig:fig2_2} clearly show that pulsars with DM>100 pc/cm$^3$, P<0.25 s and P>3 s are not detected in PUMPS. In our opinion, these sensitivity ranges are associated with increased dispersion smearing in frequency channels and with scattering (for DM>100 pc/cm$^3$), poor sampling (for P<0.25 s), and low-frequency noise in power spectra (for P>3 s). The highest sensitivity in PUMPS is achieved for pulsars with DM<100 pc/cm$^3$, having periods of 0.25<P<3s, located in the area of +21$^o$<$\delta$<+42$^o$.

At the moment, about 130 pulsars and RRAT have been discovered on LPA3, instead of the 900 expected. Thus, there is a clear shortage of new pulsars. Furthermore, some of the known pulsars are not detected in the area of the sky, where PUMPS theoretically guarantees very high sensitivity. In our opinion, there are five main factors that can explain the shortage of pulsars in PUMPS:

- the paper with the expected detection of 900 new pulsars in the LOTAAS survey was published in January 2010 (\citeauthor{vanLeeuwen2010}, \citeyear{vanLeeuwen2010}). In obtaining this estimate, the authors most likely used pulsars included in the ATNF by 2009. Therefore, when taking into account the observed deficit, all ATNF pulsars detected in PUMPS and published in papers since 2009 should be added to the selection of ``new'' pulsars. There were 96 such pulsars (Appendix~\ref{Appendix B}), of which 41 pulsars were discovered in the LPA3 survey. In addition, 45 new pulsars from the present study and about 50 discovered in RRAT should be added to the list. That is, the total number of pulsars discovered after 2009 is about 200;

- some pulsars may have flattening in the spectrum or cut-off. Estimation of the average value of the spectral index at low frequencies $\alpha$=1.4-1.6; $S\sim\nu^{-\alpha}$  (\citeauthor{Lorimer1995}, \citeyear{Lorimer1995}; \citeauthor{Bates2013}, \citeyear{Bates2013}). According to (\citeauthor{Bilous2016}, \citeyear{Bilous2016}) (Fig.~C2 in the Appendix) almost half of the 158 pulsars presented in the paper show flattening or cut-off in the spectrum in the range of 100-150 MHz. When estimating the average spectral index, we will mix pulsars with steep spectra and pulsars with flattening or cut-off in the spectra. If there is a large amount for weak pulsars with cut-off or flattening in the spectrum, then the expected flux densities from such pulsars at a frequency of 111 MHz can be highly overestimated. In this case, the sensitivity in the PUMPS survey may not be sufficient to detect some part of the new pulsars;

- sensitivity estimates in the PUMPS survey are made based on the fact that the observed pulsar flux density does not change with time. This is obviously not the case. In the paper \citeauthor{Tyulbashev2023} (\citeyear{Tyulbashev2023}) the attempt was made to estimate the integral flux densities of known pulsars based on the average profiles obtained on individual days and based on the height of harmonics obtained by summing the power spectra for 2000 observation sessions. Profiles for individual days were obtained for sessions in which the power spectra for these days showed the maximum S/N. It turned out that the average integral flux density over the entire observation period may be 5-10 times less than the flux density on the ``best'' day. Therefore, it may turn out that our sensitivity is not 10 times higher than the declared sensitivity in the LOTAAS survey (\citeauthor{Sanidas2019}, \citeyear{Sanidas2019}), as we expect (\citeauthor{Tyulbashev2022}, \citeyear{Tyulbashev2022}), but less. At the same time, it should be noted that the sensitivity in PUMPS should be higher than the sensitivity in LOTAAS, since in a blind search we found all new pulsars from the LOFAR survey that fell into the area with declinations -9$^o$<$\delta$<+42$^o$. At the same time, a number of pulsars were discovered that were not detected in LOTAAS. On the other hand, during the LOTAAS survey, raw data was recorded at hourly intervals. If the source is variable, and at the time of recording it had a minimum of radiation, for example, due to refractive scintillating on the interstellar medium, it could be lost when searching on LOFAR, but can be detected in repeated observations;

- variation of the observed properties of pulsars. In the papers \citeauthor{Young2014} (\citeyear{Young2014}); \citeauthor{Smirnova2022} (\citeyear{Smirnova2022}) the sources J1107-5907 and J0609+16 are shown, which behave part of the time as ordinary pulsars, and part of the time as RRATs. In the paper \citeauthor{Tyulbashev2018AA} (\citeyear{Tyulbashev2018AA}) it is noted that the pulsar J0302+2252 is observed as an usual pulsar on the LPA, and in RRATalog\footnote{http://astro.phys.wvu.edu/rratalog/} (\citeauthor{Deneva2016}, \citeyear{Deneva2016}) it is considered as RRAT. There is also a reverse situation. Known ATNF pulsars J1503+211 (P=3.3140 s; DM=3.26 pc/cm$^3$), J2325-0530 (P=0.8687 s; DM=14.9 pc/cm$^3$), J2346-0609 (P=1.1815 s; DM=22.5 pc/cm$^3$) were discovered on the LPA by their individual pulses\footnote{https://bsa-analytics.prao.ru/transients/pulsars/}. At the same time, with the accumulation of power spectra over 8 years, their regular radiation was not detected. This means that there is an unknown share of pulsars with transient properties, and there is also a share of pulsars that at low frequencies can be seen as RRAT, and at high frequencies as ordinary pulsars. Pulsars with intermittency and other properties causing selection are specially marked in the search for pulsars on Arecibo (\citeauthor{Deneva2009}, \citeyear{Deneva2009}). In Appendix~\ref{Appendix C}, we have placed 20 ATNF pulsars (column 7, Table~\ref{tab:tab3}) that were not detected in our search in the area  that provides maximum sensitivity in the survey (+21$^o$<$\delta$<+42$^o$; P>0.5 s; DM<200 pc/cm$^3$). These can be pulsars with strong variability, intermittent pulsars, pulsars with long nullings, with cut-off or with flattening in the spectra, as well as pulsars of the RRAT type;

- Table~\ref{tab:tab3} and Table~\ref{tab:tab4} clearly show that in the area with coordinates +21$^o$<$\delta$<+42$^o$, the percentage of pulsars lost is low. In this area, we can talk about the discovery of almost all pulsars with DM<100 pc/cm$^3$ and periods of 0.25<P<3 s. In the same area, we can also expect the detection of new pulsars at the sensitivity levels expected in PUMPS. In the area -9$^o$<$\delta$<+21$^o$, in the PUMPS survey, half of the known pulsars are lost. The reason why these losses occur does not matter, it is important that the guarantee of high sensitivity of PUMPS applies only to a relatively small area (approximately 6,500 sq.deg.), and estimates of the number of new pulsars obtained in \citeauthor{vanLeeuwen2010} (\citeyear{vanLeeuwen2010}) related to the LOTAAS survey, which took place in an area of 20,000 sq.deg. (\citeauthor{Sanidas2019}, \citeyear{Sanidas2019}).  

It is difficult to determine the share of pulsars lost and the causes of losses. Data processing using FFA is currently being prepared. For pulsars with a low duty cycle, the sensitivity during the search can increase by 5 or more times (see Fig.4 in \citeauthor{Singh2022} (\citeyear{Singh2022}). An algorithm for automatic selection of power spectra for pulsars with strong variability is also being developed. It will allow to select and sum up power spectra only for data in which a periodic signal is likely to be detected. After processing the data by two methods, we hope to increase the number of pulsars.

The histograms in Fig.~\ref{fig:fig2_2} and Table~\ref{tab:tab3} and Table~\ref{tab:tab4} clearly show that at the moment the sample of pulsars close to full is determined by the following conditions: +21$^o$<$\delta$<+42$^o$; 0.25<P<3 s; DM<100 pc/cm$^3$, |b|>10$^o$. There are 93 pulsars in the resulting complete sample (see Appendix~\ref{Appendix D}).

In conclusion, we note the main results. A search for pulsars in the area  -9$^o$<$\delta$<+55$^o$ was carried out for P>0.025 s and DM<1000 pc/cm$^3$. In a blind search using summed up power spectra, 330 pulsars were detected. Of these, 39 turned out to be new pulsars. The P and DM of new pulsars are within the boundaries of 0.1459<P<2.924 s and 13<DM<145 pc/cm$^3$. The median values of P and DM of the pulsars found are approximately at 0.9 s and 43 pc/cm$^3$. Average profiles were obtained for 6 pulsars. DM has been refined for 7 previously detected pulsars.

\section*{Acknowledgements}
The study was carried out at the expense of a grant Russian Science Foundation 22-12-00236\footnote{https://rscf.ru/project/22-12-00236/}. The authors are grateful to the antenna group for their active assistance in conducting observations and L.B. Potapova for help with the preparation of some figures. We are grateful to the anonymous referee, whose comments allowed us to improve our manuscript.

\section*{DATA AVAILABILITY} 
The PUMPS survey not finished yet. The raw data underlying this paper will be shared on reasonable request to the corresponding author. The additional figures of detected pulsars are on the website\footnote{https://bsa-analytics.prao.ru/en/pulsars/known/}.

\newpage
\appendix

\section{List of known ATNF pulsars detected in a blind search.}\label{Appendix A}

There are 291 pulsars in the list. 184 pulsars were detected in the early search as objects of regular radiation or by individual pulses. 108 known pulsars are detected in this study. The area with the coordinates $+42^o<\delta<+55^o$ has not been studied before. Pulsars that were not detected earlier in the PUMPS survey are highlighted in bold. Maps proving the discovery of pulsars can be found on the website\footnote{https://bsa-analytics.prao.ru/en/pulsars/known/}.  

J0006+1834, \textbf{J0014+4746}, J0034-0721, J0039+35, J0048+3412, J0051+0423, \textbf{J0055+5117}, \textbf{J0056+4756}, \textbf{J0058+4950}, J0109+11, J0122+1416, J0137+1654, J0146+31, J0158+21, \textbf{J0212+5222}, J0220+3626, \textbf{J0227+3356}, \textbf{J0229+20}, \textbf{J0241+16}, J0244+14, J0302+2252, J0304+1932, J0305+11, J0317+13, J0323+3944, J0332+5434, \textbf{J0335+4555}, \textbf{J0343+5312}, J0349+2340, \textbf{J0355+28}, J0358+4155, \textbf{J0358+5413}, \textbf{J0408+551}, J0417+35, J0421+3255, J0435+2749, \textbf{J0454+4529}, \textbf{J0454+5543}, J0457+23, J0517+2212, \textbf{J0518+5416}, J0525+1115, J0528+2200, J0534+2200, \textbf{J0538+2817}, J0540+3207, J0543+2329, J0546+2441, J0555+3948, \textbf{J0601-0527}, \textbf{J0608+00}, J0608+1635, \textbf{J0609+2130}, \textbf{J0611+1436}, J0611+30, J0612+37216, J0612+3721, J0613+3731, J0614+2229, \textbf{J0621+0336}, J0623+0340, \textbf{J0624-0424}, \textbf{J0627+0706}, J0629+2415, J0659+1414, \textbf{J0742+4334}, J0754+3231, \textbf{J0811+37}, J0813+22, \textbf{J0815+0939}, \textbf{J0815+4611}, J0823+0159, J0826+2637, J0837+0610, J0857+3349, J0922+0638, J0928+30, \textbf{J0935+33}, J0943+1631, J0943+2253, J0944+4106, J0946+0951, J0947+2740, J0953+0755, \textbf{J1000+08}, J1017+3011, \textbf{J1115+5030}, J1132+25, J1136+1551, J1239+2453, J1238+2152, J1239+2453, J1242+39, J1246+2253, J1303+38, J1313+0931, J1326+33, J1404+1159, \textbf{J1426+52}, J1509+5531, \textbf{J1518+4904}, J1529+40, J1532+2745, J1536+17, J1538+2345, J1543-0620, J1543+0929, J1549+2113, \textbf{J1555+01}, J1607-0032, \textbf{J1612+2008}, J1614+0737, J1627+1419, \textbf{J1628+4406}, J1635+2418, \textbf{J1635+2332}, \textbf{J1638+4005}, \textbf{J1643+1338}, J1645-0317, J1645+1012, J1649+2533, J1652+2651, J1657+3304, J1707+35, J1722+35, J1720+2150, J1728-0007, \textbf{J1735-0243}, J1735-0724, J1740+1311, J1740+27, J1741+2758, J1741+3855, J1745+1252, \textbf{J1745+42}, J1746+2540, J1746+2245, J1752+2359, J1758+3030, \textbf{J1800+5034}, \textbf{J1801-0357}, \textbf{J1810+0705}, J1813+4013, \textbf{J1814+1130}, J1814+22, J1819+1305, J1820-0427, \textbf{J1821+1715}, J1821+4147, J1823+0550, \textbf{J1825-0935}, J1832+27, J1834-0426, \textbf{J1836+51}, J1838+1650, J1841+0912, J1844+1454, J1844+41, \textbf{J1848+0647}, \textbf{J1849+0409}, \textbf{J1849+2423}, J1849+2559, J1850+1335, J1851+1259, \textbf{J1854+36}, J1900+30, \textbf{J1901+0156}, \textbf{J1904+33}, \textbf{J1906+1854}, J1907+4002, J1909+0254, \textbf{J1909+0007}, \textbf{J1909+1102}, \textbf{J1909+1859}, \textbf{J1910-0556}, J1910+0358, J1912+2525, J1913-0440, J1913+3732, \textbf{J1915+1647}, J1916+0951, J1916+3224, J1917+1353, J1917+17, \textbf{J1917+2224}, J1919+0021, J1920+2650, \textbf{J1921+1419}, J1921+1948, J1921+2153, J1922+2110, \textbf{J1923+4243}, \textbf{J1926+0431}, \textbf{J1928+28}, \textbf{J1929+00}, J1929+1844, J1929+3817, \textbf{J1931-0144}, J1932+1059, J1933+2421, \textbf{J1934+5219}, J1935+1616, \textbf{J1937+2544}, J1938+14, \textbf{J1938+0650}, \textbf{J1941+0121}, \textbf{J1941+4320}, \textbf{J1943+0609}, \textbf{J1945+07}, J1946+1805, \textbf{J1948+3540}, \textbf{J1950+3001}; J1952+1410, \textbf{J1952+3252}, \textbf{J1953+30}, J1954+2923, \textbf{J1954+3852}, \textbf{J1954+4357}, \textbf{J1955+5059}, \textbf{J2000+2920}, \textbf{J2001+4258}, J2006-0807, J2006+22, J2007+0809, J2007+0910, J2008+2513, \textbf{J2008+2755g}, \textbf{J2010+2845}, \textbf{J2016+1948}, J2017+2043, J2018+2839, J2022+21, J2022+2854, \textbf{J2022+5154}, \textbf{J2023+5037}, J2029+3744, \textbf{J2030+2228}, \textbf{J2030+55}, J2036+2835, J2037+1942, \textbf{J2040+1657}, J2043+2740, \textbf{J2045+0912}, J2046+1540, J2046-0421, J2051+1248, \textbf{J2038+35, J2047+12}, \textbf{J2053+1718}, J2055+2209, J2057+21, J2102+38, J2105+19, J2105+28, \textbf{J2108+4441}, \textbf{J2108+4516}, J2113+2754, \textbf{J2113+4644}, J2116+1414, J2122+2426, J2123+36, J2124+1407, J2139+2242, J2157+4017, J2155+2813, \textbf{J2202+21}, \textbf{J2203+50}, J2205+1444, J2208+4056, \textbf{J2208+5500}, J2209+22, J2212+2933, J2215+1538, \textbf{J2215+4524}, J2227+3038, J2234+2114, \textbf{J2235+1506}, J2253+12, J2253+1516, J2305+3100, J2306+31, J2317+2149, \textbf{J2219+4754}, \textbf{J2305+4707}, \textbf{J2313+4253}, J2333+20, \textbf{J2336-01}, \textbf{J2338+4818}, J2340+08, \textbf{J2347+02}, J2350+31, J2355+2246

\section{Pulsars discovered after 2009 and detected in PUMPS.}\label{Appendix B}

Pulsars from the area $-9^o<\delta<+42^o$. There are 98 pulsars in the list, of which 41 pulsars previously detected in PUMPS are highlighted in bold. The list does not include new pulsars from the present study.

J0039+35, J0109+11, J0122+1416, \textbf{J0146+31}, J0158+21, \textbf{J0220+3626}, J0227+3356, \textbf{J0229+20}, \textbf{J0241+16}, J0244+14, \textbf{J0302+2252}, J0305+11, J0317+13, J0349+2340, \textbf{J0355+28}, J0358+4155, \textbf{J0421+3255}, J0457+23, J0555+3948, \textbf{J0608+00}, J0608+1635, \textbf{J0611+1436}, J0612+37216,  J0613+3731, \textbf{J0621+0336}, J0623+0340, \textbf{J0811+37}, J0813+22, J0857+3349, \textbf{J0928+30}, \textbf{J0935+33}, J0944+4106, \textbf{J1000+08}, J1017+3011, \textbf{J1132+25}, J1242+39, J1303+38, J1326+33, \textbf{J1529+40}, J1536+17, J1538+2345, J1555+01, \textbf{J1612+2008}, \textbf{J1635+2332}, \textbf{J1638+4005}, \textbf{J1643+1338}, \textbf{J1657+3304}, J1707+35, J1722+35, \textbf{J1735-0243}, \textbf{J1740+27}, J1741+3855, J1745+1252, \textbf{J1810+0705}, J1814+22, J1821+4147, J1832+27, J1844+41, J1849+2559, \textbf{J1854+36}, \textbf{J1904+33}, J1910-0556, J1913+3732, J1916+3224, J1917+17, \textbf{J1917+2224}, \textbf{J1928+28}, J1929+3817, \textbf{J1931-0144}, J1938+14, \textbf{J1941+0121}, \textbf{J1945+07}, J1951+3001, \textbf{J1953+30}, \textbf{J1954+3852}, \textbf{J2000+2920}, J2006+22, \textbf{J2008+2755g}, \textbf{J2010+2845}, J2022+21, J2036+2835, J2051+1248, \textbf{J2053+1718}, J2057+21, J2105+19, J2105+28, J2122+2426, J2123+36,  J2208+4056, J2209+22, J2253+12, J2306+31, J2333+20, \textbf{J2336-01}, J2340+08, \textbf{J2347+02}, J2350+31, \textbf{J2355+2246}

\section{ATNF pulsars not detected in PUMPS.}\label{Appendix C}

Pulsars located in the area $+21^o<\delta<+42^o$ are selected,where there is the minimum level of interference observed. Periods and dispersion measures of these pulsars(P>0.5 s, DM<200 pc/cm$^3$) are such that sensitivity losses due to dispersion smearing and scattering are minimal \citep{Tyulbashev2022}. When searching at an interval of 8 years for these pulsars, the sensitivity is no worse than 0.5 mJy at a frequency of 111 MHz.

J0414+31, J0658+2936, J0927+2345, J1629+33, J1822+2617, J1829+25, J1912+2104, J1919+2621, J1929+2121, J1931+30, J2011+3006g, J2015+2524, J2017+2819g, J2037+3621, J2044+28, J2111+2106, J2129+4119, J2145+21, J2151+2315, J2307+2225

\section{A complete sample of pulsars at the frequency 111 MHz.}\label{Appendix D}

Complete sample for pulsars located in the extragalactic plane: $-9^o<\delta<+42^o$; 0.25<P<3 s; DM<100 pc/cm$^3$; |b|>10$^o$. Starting from $S_{int}$=0.5 mJy, PUMPS should detect all pulsars (see Table~\ref{tab:tab2}; \citet{Tyulbashev2022}). Pulsars opened in PUMPS in different years are highlighted in bold.

J0039+35, J0048+3412, \textbf{J0129+36}, \textbf{J0146+31}, \textbf{J0149+29}, J0156+3949, J0158+21, \textbf{J0220+3626}, J0227+3356, \textbf{J0303+2248}, J0349+2340, \textbf{J0350+2341}, J0417+35, \textbf{J0421+3240}, J0435+2749, J0457+23, J0754+3231, \textbf{J0811+37}, J0813+22, J0826+2637, \textbf{J0928+3037}, \textbf{J0935+33}, J0943+2253, J0944+4106, J0947+2740, \textbf{J1003+41}, J1017+3011, \textbf{J1105+37},  J1238+2152, J1239+2453, \textbf{J1242+3938}, J1246+2253, J1303+38, \textbf{J1529+40}, J1532+2745, J1549+2113, \textbf{J1555+01} \textbf{J1635+2332}, J1635+2418, \textbf{J1638+4005}, J1649+2533, J1652+2651, \textbf{J1657+3248}, J1720+2150, \textbf{J1721+3524}, \textbf{J1740+2728}, J1741+2758, J1741+3855, J1746+2540, J1752+2359, J1758+3030, J1813+4013, J1814+22, J1821+4147, \textbf{J1832+27}, \textbf{J1844+21}, \textbf{J1844+41}, \textbf{J1845+2147}, \textbf{J1848+26}, J1849+2423, J1849+2559, \textbf{J1854+40}, J1854+36, J1900+30, J1904+33, J1907+4002, J1913+3732, \textbf{J1916+38}, \textbf{2047+12}, J2055+2209, J2057+21, J2105+28, J2113+2754, J2122+2426, \textbf{J2122+3432}, J2139+2242, J2155+2813,  J2157+4017, \textbf{J2201+33}, \textbf{J2203+21}, J2207+4056, \textbf{J2209+22}, \textbf{J2210+2117}, J2212+2933,  J2227+3038, \textbf{J2229+40}, J2234+2114, J2305+3100, J2306+31, \textbf{J2312+21}, J2317+2149, \textbf{J2350+31}, J2355+2246

\bsp	
\label{lastpage}
\end{document}